 \documentclass[final,5p,times,twocolumn,numbers]{elsarticle}

\usepackage{amsmath,amssymb,amsfonts}
\usepackage{algorithmic}
\usepackage{graphicx}
\usepackage{textcomp}
\usepackage{booktabs}
\usepackage{url}
\usepackage[utf8]{inputenc}
\usepackage{float} 

\usepackage{lipsum}

\journal{}

\begin{document}

\begin{frontmatter}



\title{Semi-Unsupervised Microscopy Segmentation with Fuzzy Logic and Spatial Statistics for Cross-Domain Analysis Using a GUI}

\author[1]{Surajit Das \corref{cor1}}
\ead{mr.surajitdas@gmail.com}
\cortext[cor1]{Corresponding author}

\author[1]{Pavel Zun}
\ead{pavel.zun@gmail.com} 

\address[1]{ITMO University, St. Petersburg, Russia}

\begin{abstract}
Brightfield microscopy of unstained live cells is challenging due to low contrast, dynamic morphology, uneven illumination, and lack of labels. Deep learning achieved SOTA performance on stained, high-contrast images but needs large labeled datasets, expensive hardware, and fails under uneven illumination. This study presents a low-cost, lightweight, annotation-free segmentation method by introducing one-time calibration-assisted unsupervised framework adaptable across imaging modalities and image type. The framework determines background via spatial standard deviation from the local mean. Uncertain pixels are resolved using fuzzy logic, cumulative squared shift of nodal intensity, statistical features, followed by post-segmentation denoising calibration which is saved as a profile for reuse until noise pattern or object type substantially change. The program runs as a script or graphical interface for non-programmers. The method was rigorously evaluated using \textit{IoU}, \textit{F1-score}, and other metrics, with statistical significance confirmed via Wilcoxon signed-rank tests. On unstained brightfield myoblast (C2C12) images, it outperformed \textit{Cellpose 3.0} and \textit{StarDist}, improving IoU by up to 48\% (average IoU = 0.43, F1 = 0.60). In phase-contrast microscopy, it achieved a mean IoU of 0.69 and an F1-score of 0.81 on the \textit{LIVECell} dataset ($n = 3178$), with substantial expert agreement ($\kappa > 0.75$) confirming cross-modality robustness. Successful segmentation of laser-affected polymer surfaces further confirmed cross-domain robustness. By introducing the \textit{Homogeneous Image Plane} concept, this work provides a new theoretical foundation for training-free, annotation-free segmentation. The framework operates efficiently on CPU, avoids cell staining, and is practical for live-cell imaging and biomedical applications. Code and data are available for reproducibility.*

\end{abstract}

\tnotetext[code]{\url{https://github.com/anonymusUID/01cvLiveCell01p}}




\end{frontmatter}




\section{Introduction}
\label{introduction}

The use of bright-field microscopy is ubiquitous in research and clinical diagnostics  for observing specimens, routine analysis with different objectives \cite{Murphy2012}. The study of biological samples, accompanied by machine learning and data science \cite{Moen2019}, often suffers from significant challenges for studying unstained (naturally pigmented) specimen which appear as low-contrast \& noisy objects under microscope.   

Deep learning (DL) tools (e.g.Cellpose 3.0) \cite{Stringer2021} achieve impressive results for stained (high contrast or pigmented) labeled data, but they consistently fail under the low-contrast conditions characteristic of unstained samples due to their dependency on training data and sensitivity to noise. In addition, scarcity of sufficient annotated data hinders the DL training process. Therefore, segmenting unstained live cells remains a significant challenge involving faster, annotation-free and training-free analysis of live cells without killing, staining, or fixing them. 

Our work bridges this critical gap by introducing a novel unsupervised methodology for live unstained cell culture images. The proposed methodology requires no annotations or pretraining while outperforming state-of-the-art (SOTA) approaches in challenging bright-field microscopy conditions.

The fundamental challenge in unstained bright-field microscopy lies in the weak contrast between cells and background, arising from subtle differences in light absorption and scattering \cite{Murphy2012}. Recent DL-based segmentation tools \cite{Stringer2021, Cutler2022} typically rely on strong contrast, limiting their effectiveness for live-cell studies where staining is impractical. This limitation becomes particularly acute when examining delicate specimens like primary cell cultures or patient-derived samples.

\subsection*{Challenges in Current Methods}
Our analysis identifies that using ML/DL faces ten core challenges of existing approaches: \textbf{(I) Low contrast \& noise –} Current SOTA models achieve IoU scores below 0.3 on unstained data (see §V), as they cannot distinguish cells from background noise without strong contrast cues. \textbf{(II) Uneven illumination –} Inconsistent brightness across the field of view (FOV) misleads ML/DL models. \textbf{(III) Optical aberrations –} Dust, air bubbles, glare, and slide imperfections introduce artifacts. \textbf{(IV) Overlapping dynamic structures –} Cells and tissues often overlap with temporal changes in shape which complicates segmentation. \textbf{(V) Annotation dependency –} Microscopy data
demands accurate annotation, requires expert knowledge, and  is both error-prone and time-consuming. Another problem is scarcity of plenty of data. \textbf{(VI) Data imbalance –} Rare biological events (e.g., abnormal cells) lead to biased models that perform poorly on rare but critical features. \textbf{(VII) Lack of standardization –} Models trained on one dataset often fail to generalize to unseen datasets or imaging conditions
due to variations in microscopes, lighting, and deviation from the sample preparation. \textbf{(VIII) 2D limitations –} Bright-field microscopy lacks depth information, obscuring 3D features. \textbf{(IX) 
Training dependency –} Supervised tools require extensive annotations \cite{Moen2019}, making them impractical for many live-cell applications. \textbf{(X) Generalization gaps –} Models trained on one dataset often fail under new imaging conditions \cite{HOQUE2024101997}, especially with varying illumination or optical setups. \\

Traditional solutions like spatial masking \cite{Netravali1977} or spectral filtering \cite{Albanwan2018} either require physical modifications to the imaging system or fail to adapt to biological variability. Computational approaches like region growing \cite{Tang2010} depend heavily on seed selection and distance metrics, making them unreliable for heterogeneous samples.

\subsection*{Our Approach}
We present a fundamentally new framework that is found to be robust under the following assumptions: I) the environment involving optical conditions and the imaging system of bright-field microscopy is set up to achieve a nearly homogeneous image plane (defined in  section 2); II) one of two image sets (either background or specimens) must have optical heterogeneity caused by its inherent texture. The propsed model combines:

\begin{itemize}
    \item \textbf{Spatial Standard Deviation from Local Mean (SSDLM)}: A novel statistical metric that works for measuring the homogeneity of the image set (Theorem 1, §II.B).
    
    \item \textbf{Fuzzy logic system}: GUI-controlled, facilitated dynamic system to resolve classification ambiguity.

    \item \textbf{Cumulative Squared Shift of Nodal Intensity (CSSNI)}: A novel spatial statistic that quantifies local intensity variations more robustly than traditional gradient-based measures in low-contrast regimes (§II.C). 
    
    \item \textbf{Adjusted variogram analysis}: Robust to uneven illumination optical artifacts.
\end{itemize}

\subsection*{Contributions}

This work introduces the first fully unsupervised segmentation framework that consistently outperforms state-of-the-art (SOTA) models from 2023–2024 on unstained bright-field microscopy images. Our method achieves a 48\% improvement in Intersection over Union (IoU) compared to Cellpose 3.0 (Table~3), with statistically significant gains ($p < 0.01$, Wilcoxon signed-rank test), all while operating on standard CPU hardware.

The core contributions of this study are summarized as follows:

\begin{itemize}
    \item \textbf{Unsupervised segmentation for live cells:} Our method eliminates the need for staining or labeling, preserving cell viability and avoiding cytotoxic effects—critical for live-cell and longitudinal studies.
    
    \item \textbf{Practicality and accessibility:} The algorithm requires no training, annotations, or GPU resources, making it deployable in resource-limited biological labs and usable by non-programmers.
    
    \item \textbf{Biological relevance:} Working directly on label-free images, the method enables morphology analysis without experimental artifacts, capturing cells in their native physiological states.
    
    \item \textbf{Software and reproducibility:} The full implementation is released as open-source software with an intuitive, GUI-based interface tailored for biomedical researchers.
    
    \item \textbf{Real-world validation:} The framework is validated on challenging datasets, including primary cell cultures and clinical microscopy samples, demonstrating robust performance in diverse imaging conditions.
\end{itemize}

\section{Related Work}
In this section, a comprehensive overview is captured based on the contemporary research that introduces unsupervised learning based on several algorithmic \& computational themes. Some notable research is included to depict the general evolutions of unsupervised learning for image segmentation problems.   Unlike the images derived from the macro imaging system (which indicates the standard imaging system) or the images derived from the remote sensing system, the microscopic images have very high spatial resolution. In some microscopy modalities, specific contrast and brightness (phase contrast, fluorescence) are also important. The imagery often adopts special types of noises and other challenges which are already discussed in the introduction, and these noises or challenges are not present in the images obtained from standard or remote sensing systems. However, this literature review is not only restricted within the scope of unsupervised learning in the field of microscopy images. It examines the overall recent advancements and accompanying gaps in unsupervised image segmentation techniques, as the methodologies developed for unsupervised learning in segmentation problems are limited.
\subsection{Unsupervised Learning for Image Segmentation Based on CNN}
Among the several studies on unsupervised learning for image segmentation, one notable method is the convolutional neural network (CNN)-based algorithm for unsupervised image segmentation \cite{Kim2020}. This approach optimizes feature extraction and clustering functions jointly, predicting cluster labels through differentiable functions. A spatial continuity loss enhances segmentation quality, while batch normalization normalizes response maps. 
PASCAL VOC 2012 and BSD500 are the used datasets for the experiment.  However, the paper exhibits some research gaps as it does not address real-time segmentation performance, especially in the case of low-contrast noisy dataset which can be comparable to the scenario of microscopy images. Additionally, the paper has not argued scientifically about the diversity of input data and about the scalability of the proposed method.
\subsection{Bayesian Statistical Model}
An old study proposes the Bayesian unsupervised satellite image segmentation method based on stochastic estimation maximization (SEM) algorithm over global methods like MAP or MPM \cite{Masson1993}. This study assesses spectral and spatial context contributions to image parameters. However, a serious research gap is discovered as the dependence on initialization affects solution reliability. Also, in certain cases, the limited exploration of spatial context contributions is observed. In case of low signal-to-noise ratio scenarios, the model needs to be developed for a good result and it should be validated for various image parameters. Another study, Image Segmentation with Adaptive Spatial Regularisation (ASR), introduced a Bayesian computation methodology accompanied by Potts-Markov random fields (MRFs) \cite{Pereyra2017}. The method marginalized regularization parameters and considered small-variance asymptotic analysis. Despite achieving comparable results to supervised approaches, it lacked exploration of alternative regularization techniques and scalability assessments. Also, Independent pixel consideration neglects neighbouring pixel influence. 

\subsection{Soft Computing Based Approaches}
A research involving Evolutionary Algorithm-Based Fuzzy Clustering (EABFC) introduces an unsupervised fuzzy clustering approach for image segmentation, combining an evolutionary algorithm (EA) with fuzzy clustering to leverage both local and non-local spatial information \cite{Mukhopadhyay2009}. The method employs a multi-objective evolutionary sampling strategy to optimize pixel selection while preserving image details, followed by label correction using entropy and spatial constraints. However, the study has two key limitations: (1) it lacks experimental validation on diverse datasets, raising concerns about generalizability, and (2) it does not support user-defined parameters, limiting customization for different segmentation tasks.
A novel approach combining fuzzy logic with Markov random field (MRF) has been proposed for image segmentation \cite{Nguyen2013}. This method develops an adaptive fuzzy inference system and utilises spatial constraints effectively. The approach is notable for implementing a new clique potential MRF function.
Fuzzy logic has been widely used for unsupervised segmentation. Fuzzy Random Fields and Unsupervised Image Segmentation proposed a fuzzy statistical model incorporating Gibbs sampling and stochastic estimation maximization (SEM) methods \cite{Caillol1993a}. The approach demonstrated robust segmentation by integrating fuzzy components into traditional statistical models. However, the study did not explore real-world applications extensively or compare with advanced segmentation techniques.
Another notable study, Estimation of Fuzzy Gaussian Mixture and Unsupervised Statistical Image Segmentation, applied adaptive iterative conditional estimation (ICE) to improve segmentation efficiency \cite{Caillol1997}. The model generalized statistical fuzzy segmentation and adapted it to contextual settings using SEM, ICE, and Expectation-Maximization (EM) algorithms. Future work aims to integrate the approach with existing segmentation techniques.

\subsection{Unsupervised Domain Adaptation for Microscopy Images}
Panoptic Domain Adaptive Mask (PDAM), based on Domain Adaptive Mask R-CNN (DAM), offers a novel segmentation strategy but struggles with domain shift due to contextual discrepancies \cite{Liu2021}. This approach works with R-CNN and uses cycleGAN with an auxiliary objects inpainting mechanism. The former is responsible for synthesising images alike to the target, while the later one is responsible for reinforcing the image construction. The method encounters a problem regarding domain shift due to contextual information discrepancies and exhibits inadequate adaptation in the feature level for large domain gaps. 
Another approach puts forward an encoder-decoder-based multi-task learning model to cluster pixels according to foreground, background and cell boundaries as unsupervised domain adaptation. This method requires further improvements in domain-regularising cost functions and performance metrics, as the Dice metric is insensitive to clustered cell isolation. Also, the article does not discuss fixing the issue of ad-hoc parameter estimation \cite{Mukherjee2023}.

\subsection{Hyperspectral Image Segmentation}
In another approach, hyperspectral microscopy image segmentation combines both unsupervised deep learning (UHRED) for denoising and supervised deep learning (SHRED) for enhancement followed by K-means clustering and mean squared error for loss calculation \cite{Abdolghader2021}. The method uses the Adam optimizer for determining the parameters of the model. However, challenges pertaining to overlapping species classification and automation of saturated pixel identification remain unanswered.

\subsection{Classical and Hybrid Approaches}
The traditional technique along with machine learning has tried to accomplish the unsupervised learning, competitive learning, fuzzy c-means clustering, and Gibbs random fields to improve tissue component segmentation through an iterative conditional modes (ICM) algorithm adaptation. The prominent lacks of this approach lie in limited accuracy produced, subjectivity in setting thresholds and the need for ICM algorithm adaptation \cite{Gaddipati1997}.
Another research proposed a classical approach involving edge detection and morphological processes, called MPS-Based Image Segmentation for Bright-field Microscopy \cite{Cepa2018}. The method has been implemented in open-source software Fiji. Histogram equalisation, edge detection by the Canny edge detector and filling holes by using a maximum filter are the key steps to segment the total cell area by creating a binary image. Though it works with various cell types, it has a serious lack of handling the scenarios where the cell border is obscured. Also, the report about huge testing and comparison outcomes across diverse scenarios is absent. It requires better parameter standardisation across microscopy setups. Another work, \textbf{Self-Supervised Learning (SSL \footnote{ abbreviation replicated from publisher, "Nature  Communication". \url{https://www.nature.com/articles/s42003-025-08190-w}}) Approaches} \cite{lam2025ssl}, advances annotation-free segmentation via optical flow-based pseudo-labeling, achieving $F_1$ scores of 0.77--0.88 on fluorescence images (as per the author's claim). However, for bright-field myoblasts, it exhibits major limitations: (1) 50--60 s per image due to iterative optical flow, and (2) failure on 60\% of low-contrast samples where texture features are unreliable. This hampers longitudinal studies needing speed and consistency. The error "Either no cells found or all cells are touching the border" arises when cells are undetected or too close to image edges.

\section{Methodological Background \& Technical Discussion}

\subsection{Introducing Image Plane}
\textbf{Image Plane:} A theoretical and physical 2$D$ space, which is a collection of pixels that are arranged in an array and have intensity values $I$ is called an image plane \cite{Born1999,Jaffe2001, Goodman2005}. The intensity value for any pixel is determined as the outcomes of transmittance or reflectance function $T$, the illumination factor is governed by the physical plane of the sensor $L$ and optical distortions/aberrations or sensor non-uniformity $\Delta$ \cite{Gibson1991, Goodman2005} and hence $I$ encodes a convolution of optical, spatial, and noise characteristics that must be considered for accurate segmentation or enhancement \cite{Zhang2017}. Let \( I'(x', y') \) be the intensity function defined as:
\begin{equation}
I=I'(x', y') = T\left(P^{-1}(x', y', 1)^\top\right) \cdot L(x', y') + \Delta(x', y'),
\label{eq:intensity_function}
\end{equation}
where: $I'(x', y')$ is the intensity function at coordinates $(x', y')$, $T$ is a 3D property (e.g., texture or reflectance), $P^{-1}$ is the inverse projection matrix mapping 2D image coordinates to 3D space, $L(x', y')$ is a 2D modulation term (e.g., lighting), $\Delta(x', y')$ represents noise or offsets, $I$  is the scalar intensity value at a specific point $(x', y')$.

\subsection{Notion of Homogeneous Image Plane \& Homogeneity}
\subsubsection{Homogeneous Image Plane}
According to the hypothesis considered, in our context of microscopy, a homogeneous image plane is defined as an image plane where the equation (1) can be written as:
\begin{equation}
\begin{split}
I'(x',y') = T(P^{-1}(x',y',1)^T) \cdot L_0 
\end{split}
\end{equation}
Here, \( L_0 \) is a constant (or nearly constant) representing uniform illumination across the field of view (FOV). Any observed intensity variation arises solely from the object's texture, not from inconsistencies in the optical system, illumination, or imaging medium. Nullifying the \( \Delta \) terms corresponds to an idealization of the optical properties---i.e., aberrations, distortions, or other system-induced inconsistencies (such as those introduced by lenses or other optical elements) are assumed to be zero or negligible.  

According to our hypothesis in microscopy, the background of an image should be homogeneous, as it lacks inherent texture. However, in reality, achieving a perfectly homogeneous plane is impossible, since neither \( \Delta \) equals zero nor \( L \) remains strictly constant. Therefore, we introduce a new computationally efficient metric (compared to traditional homogeneity measures based on the gray-level co-occurrence matrix (GLCM)) to measure homogeneity and establish a threshold for it.

\subsubsection{Homogeneity of Region of Interest (ROI) in Image} 
Traditionally, homogeneity is measured with the help of gray level co-occurrence matrix (GLCM) \cite{Singh2017} and formulated as:  
\[ 
  H_d^{\theta}=\sum_{i=1}^n \sum_{j=1}^n \frac{p_d^{\theta}(i,j)}{1+|i-j|}
\]
where $n$ is the total number of gray levels in the neighbourhood selected, $  p_d^{\theta}(i,j)$ denotes the probability of a pixel pair having intensities $i$ and $j$ respectively at a certain $d$ distance and angle $\theta$. Calculating homogeneity is computationally expensive \cite{Singh2017}. \\ \\
 We measure mean and standard deviation of spatial standard deviation with respect to localized mean (SD of SSDLM) for a neighbourhood.  Unlike the homogeneity measures in the GLCM, this technique has limitations in capturing the homogeneity along every sense of direction, but it is  proved below that the SSDLM inversely varies with the metric homogeneity (if it exists) along any sense of direction.\\ \\
Let us consider a $3 \times 3$ neighbourhood around any pixel intensity $\omega_{(i,j)}$. In this case, $i$ and $j$ represent the cartesian coordinate of that pixel and $\omega$ is a function that maps to the intensity of that pixel. 
$$\begin{pmatrix}
\omega_{(i-1,j-1)} & \omega_{(i-1,j)} & \omega_{(i-1,j+1)}\\
\omega_{(i,j-1)} & \omega_{(i,j)} & \omega_{(i,j+1)} \\
\omega_{(i+1,j-1)} & \omega_{(i+1,j)} & \omega_{(i+1,j+1)} 
\end{pmatrix}$$ 
Therefore,  considering $\overline{\omega}$ as the mean value of $\omega$, SSDLM for the neighbourhood of $\omega_{i,j}$ will be represented as:
\begin{equation}
SSDLM_{(i,j)}= \sqrt{\sum_i \sum_j \frac {(\omega_{(i,j)}.-\overline{\omega})^2}{N}}  
\end{equation}

From the context of GLCM matrix, we know that the homogeneity directly varies with the contrast \cite{Singh2017}. Hence,
$$\sum_{i=1}^n \sum_{j=1}^n \frac{p(i,j)}{1+|i-j|} \propto
\frac{1}{ \sum_{i=1}^{n} \sum_{j=1}^{n} (i - j)^2 \cdot p(i, j)} 
$$

$$
Again, \sum_{i=1}^{n} \sum_{j=1}^{n} (i - j)^2 \cdot p(i, j) \propto \sqrt{\sum_i \sum_j \frac {(\omega_{(i,j)}.-\overline{\omega})^2}{N}} $$ \text{[ according to the rules of variation]}

Therefore, $SSDLM_{(i,j))}$ is inversely proportional to homogeneity. 

In order to determine the threshold of homogeneity (lower bound), background sampling was performed from a large population of images (10000 images). Random background pixels were then selected, and using a fixed kernel size $(5)$, the SSDLM was calculated. The mean and standard deviation (SD) of SSDLM were computed, and the threshold was set to $3 \times$ SD in the positive scale, which was used as the SSDLM (homogeneity) threshold (lower bound) for a given modality. For bright-field microscopy, the homogeneity threshold was set to $4.23$. This value may vary depending on the modality or surface; for example, for phase-contrast microscopy, it is $3.5$, and for fiber sheets, it is approximately in the same range. However, this value can be adjusted by the user through tuning the corresponding hyperparameter.

\subsection{Other Spatial Statistical Metric Used}. 
\subsubsection{Moran’s I} Moran’s I is a measure of spatial autocorrelation, which is defined as:
$$I = \frac{n \sum_{i}\sum_{j}w_{ij}(x_i - \bar{x})(x_j - \bar{x})}{\sum_{i}\sum_{j}w_{ij} \cdot \sum_{i}(x_i - \bar{x})^2} $$ 
$n:$ Number of spatial units, $x_i$, $x_j$: Values at locations i and j, where, $i$ and $j$ are the labels of two locations, $\bar{x}$: Mean of all values, $w_{ij}$: Spatial weight between locations i and j (e.g., inverse distance).
$I > 0$: Positive autocorrelation (clustering), $I < 0$: Negative autocorrelation (dispersion) and 
$I = 0$: No autocorrelation (randomness)

\subsubsection{Cumulative Squared Shift of Nodal Intensity (CSSNI)}
CSSNI quantifies local intensity variations in greyscale images by computing the sum of squared intensity differences (SSID) between each pixel and its 8-connected neighbour. The mathematical formulation of this metric is given below:\\
Let $\omega_{(i, j)} $ denote the intensity value of a pixel at the location $(i, j)$ in a patch size of $ ( M \times N )$ in a greyscale image. The CSSNI measure is defined as:
\begin{equation}
CSSNI = \frac{1}{2} \sum_{i=1}^{M} \sum_{j=1}^{N} \sum_{(m,n) \in \mathcal{N}_{(i,j)}} (\omega_{(i, j)} - \omega_{(m, n)})^2
\end{equation}
where \( \mathcal{N}_{(i,j)} \) represents the set of 8-connected neighbouring pixels. The division by 2 prevents double-counting since each pixel pair contributes to the sum twice.

\subsubsection{Adjusted Variogram}

Adjusted variogram is a variogram-like measure for spatial intensity variation. The spatial variation of the intensities of the pixels within an image patch is commonly analyzed using the semivariogram \cite{Oliver2010}, which quantifies the relationship between the differences in the intensity of the pixels and their spatial separation. A conventional empirical semivariogram \( \gamma(h) \) is defined as:

\[
\gamma(h) = \frac{1}{2N(h)} \sum_{i=1}^{N(h)} (x_i - x_j)^2
\]

Where  $h$  is the spatial lag,  $x_i$ and  $x_j$  are there pixel intensities at locations  $i$  and $j$ (which are two different labels of the location), and  $N(h)$  is the number of pixel pairs separated by \( h \). Standard variograms rely on binning pixel pairs based on discrete lags to estimate spatial dependence.

This study introduces adjusted variogram providing a global estimate of intensity variation without explicit lag binning. The function computes the average squared intensity differences between all pixel pairs, normalised by their Euclidean distance in a specified neighbourhood. It can be formulated as:

\begin{equation}
\gamma(h) = \frac{1}{2} \cdot \mathbb{E} \left[ \frac{(z(x) - z(x+h))^2}{d(x, x+h)} \right]
\end{equation}

Where:
\begin{itemize}
  \item \( z(x) \) is the pixel intensity at position \( x \),
  \item \( z(x+h) \) is the pixel intensity at position \( x+h \),
  \item \( d(x, x+h) \) is the Euclidean distance between positions \( x \) and \( x+h \),
  \item \( h \) is the spatial distance between pixels.
\end{itemize}

This method differs from conventional variograms in three key aspects:
\begin{enumerate}
    \item \textbf{No explicit lag binning}: Instead of computing semivariance for specific lag distances, this approach aggregates all pairwise intensity differences.
    \item \textbf{Global averaging}: The function provides a single scalar estimate of spatial intensity variation rather than a curve over multiple lags.
    \item \textbf{1D transformation}: The patch is flattened into a sequence, which simplifies computation but does not fully preserve 2D spatial relationships.
\end{enumerate}

The adjusted measure does not replace a traditional variogram; it serves as a computationally efficient alternative for capturing global intensity variation, here used with a goal of texture analysis and spatial feature extraction. Its overall advantages are i) fast and lightweight, ii) single-scalar output and iii) good for local analysis. 

\subsection{Computational Complexity Analysis}

In this section, we analyze the computational complexity of the proposed methods to determine their efficiency in processing microscopic images.

\subsubsection{Time Complexity of Functions}

We analyze the time complexity of each function used in the algorithm. Here we denote the total number of pixels in the image as $N=n \times m$   and the patch size as \( N_p \) (typically 5x5, 7x7 or 9x9).

\begin{itemize}
    \item \textbf{Calculating Moran's I for a neighbourhood):} The dominant terms arise from the distance matrix computation, weight matrix construction and dilation. Hence, the overall time complexity of the function is: $ \mathcal{O}(N^2) = \mathcal{O}((n \times m)^2)$

    \item \textbf{Calculating Adjusted Variogram-like Measure:} This function computes pairwise distances and squared differences in \( O(N_p^2) \), leading to an overall complexity of \( O(N_p^2) \), where \( N_p \) is the number of pixels in the patch.
    
    \item \textbf{Calculating Cumulative Squared Shift of Nodal Intensity (CSSNI):} This iterates over all pixels and their 8-connected neighbour, resulting in a time complexity of \( O(N) \).
    
    \item \textbf{Fuzzy Module:} This function performs basic arithmetic and comparisons, leading to a constant time complexity of \( O(1) \).
\end{itemize}

\subsubsection{Overall Time Complexity}

The most computationally expensive functions are calculating adjusted variogram and calculating Moran's I, which run in  $O(N_p^2)$  per pixel ($p \le 11$). Since these functions are not applied to every pixel in the image (used only where fuzzy cannot identify the pixels for the foreground or background set), the overall complexity of the algorithm will be much less than:    $O(N \cdot N_p^2).$
Given that,  $N_p$  is relatively small, the algorithm remains efficient for large images.

This analysis confirms that the proposed method is computationally feasible for high-resolution microscopic images while maintaining accuracy in segmentation and classification. However, Moran’s I can be accelerated using quadtree-based spatial indexing, Variogram calculations can be optimized with multi-resolution approximations. Also, parallel processing (Graphical Processing Unit (GPU) acceleration) can significantly reduce runtime for large images.

\section{Methodology} 
\begin{figure}[htbptbp!]
  \centering
  \includegraphics[width=.48\textwidth]{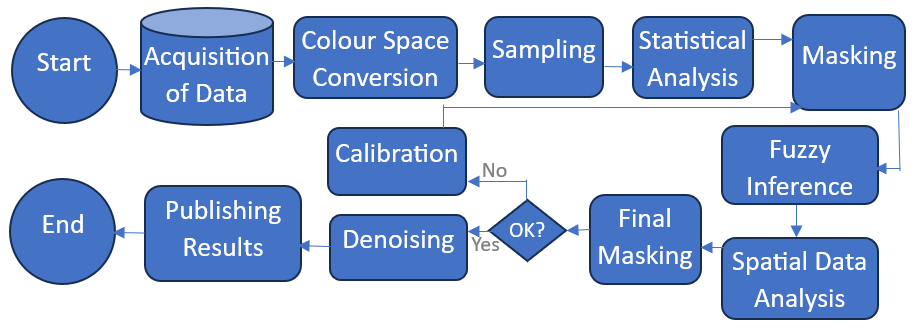}
  \caption{schematic diagram of end-to-end workflow}
  \label{fig:Workflow}
\end{figure}

The schematic diagram for end-to-end workflow is attached (Fig.~\ref{fig:Workflow}). All the steps of the workflow are described in subsequent sections. 

\subsection{Data Acquisition} The first experiments used 10 primary real-time $1920 \times 1440$ images (\textbf{Dataset-1}\footnote{Dataset avilable on github link.}) of low-contrast myoblast (C2C12 cell) cultures from our lab, featuring irregular, unstained cells with often obscured boundaries due to motion. The cells were phographed using bright-field microscope. The methodology was also tested on the publicly avaiable LIVECell dataset (\textbf{Dataset-2}\footnote{\url{https://sartorius-research.github.io/LIVECell/}}) \cite{Edlund2021}, which comprises diverse cell types and 3,180 cells, with a modality of phase-contrast microscopy. Finally, we segment the images (\textbf{Dataset-3}\footnote{Dataset avilable on github link.}) of controllable laser traces on surface patterns and evaluate the robustness of our model across domains.

\subsection{Colour Space Conversion} Most libraries (e.g., OpenCV) convert RGB to grayscale using a perceptual luminance model based on human vision sensitivity: $\text{Gray}=0.299R+0.587G+0.114B$. In contrast, this study applies an \textbf{average-based conversion}: $\text{Gray}=\frac{R+G+B}{3}$, assigning \textbf{equal weights} to all channels to ensure unbiased sensitivity in the treatment of second-order exponents of intensity variations across RGB components.

\subsection{Primary Masking} In this step, the image background is considered a nearly homogeneous, monochromatic medium, and the $SSDLM_{\omega}$  of the patches of all the pixels are checked. This assumption is grounded in the lighting mechanism of Bright-Field microscopy; however, errors / uncertainties arising from several factors will be handled in subsequent steps. For a pixel and its patch, if its $SSDLM_{\omega}$ is found to be lower than the lower bound, the pixel is turned into black by replacing the pixel value with zero.

\subsection{Fuzzy Inference System}
Next, intensity transformation using fuzzy logic is performed, with sub-steps detailed below.

\subsubsection{Fuzzy Membership Function Definitions}
\begin{figure}[htbp!]
  \centering
  \includegraphics[width=.48\textwidth]{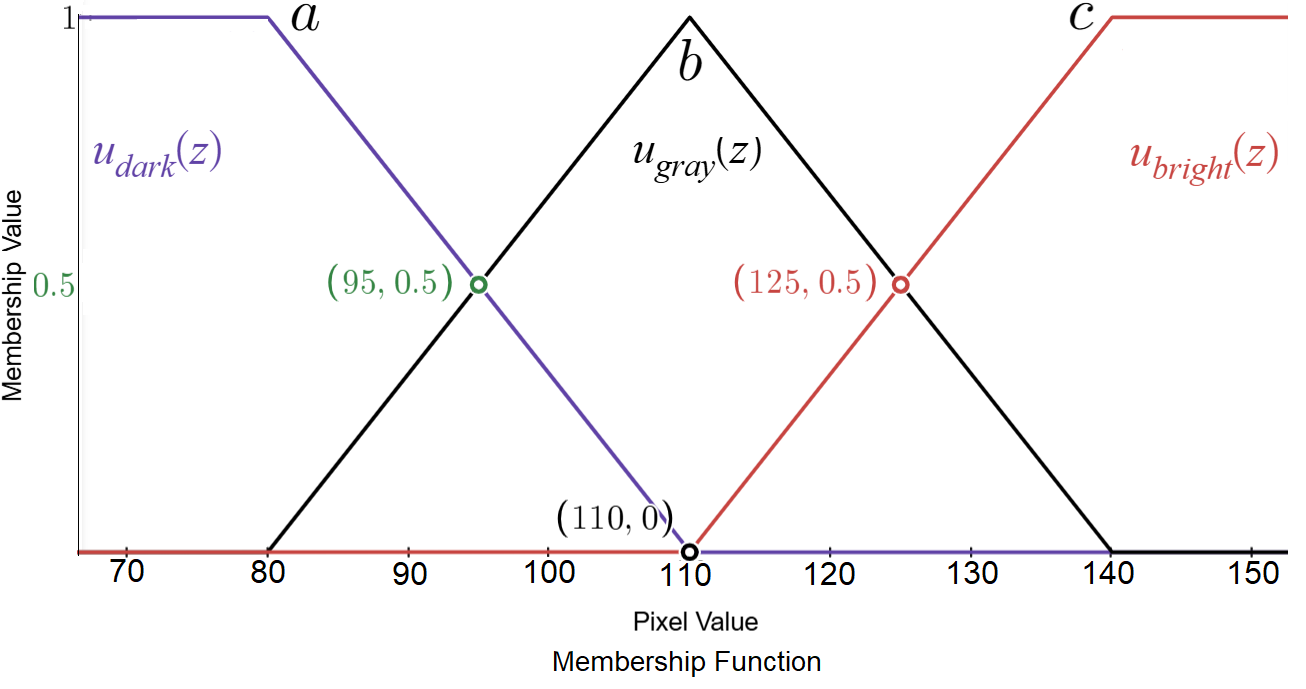}
  \caption{Membership functions $\mu_{dark}$, $\mu_{gray}$, and $\mu_{bright}$ for black, gray, and white regions.}
  \label{fig:Fuzzy_Membership_func}
\end{figure}

Three membership functions categorize pixel intensity, with parameters $\alpha=\beta = b =110$, $c=140$, and $a=80$ (Fig.~\ref{fig:Fuzzy_Membership_func}).

\paragraph{Half-Trapezoidal Decreasing} $u_d(x)$
\begin{equation}
    u_d(I_{\text{eff}}) =
    \begin{cases}
    1, & I_{\text{eff}} \leq a \\
    \max\left(\frac{\alpha - I_{\text{eff}}}{\alpha - a}, 0\right), & \text{otherwise}
    \end{cases}
\end{equation}

\paragraph{Triangular} $u_g(x)$
\begin{equation}
    u_g(I_{\text{eff}}) = \max\left( \min\left( \frac{I_{\text{eff}} - a}{b - a}, \frac{c - I_{\text{eff}}}{c - b} \right), 0 \right)
\end{equation}

\paragraph{Half-Trapezoidal Increasing} $u_b(x)$
\begin{equation}
    u_b(I_{\text{eff}}) =
    \begin{cases}
    1, & I_{\text{eff}} \geq c \\
    \max\left( \frac{I_{\text{eff}} - \beta}{c - \beta}, 0 \right), & \text{otherwise}
    \end{cases}
\end{equation}

\subsubsection{Hyperparameter Tuning}
Default values are $\alpha=\beta=110$, $b=110$, $c=140$, $a=80$. These cannot be directly modified via the calibration window (Fig.~4), but the “Shift Gray” slider adjusts $b$ (default 110), shifting the gray midpoint. The “Span Gray” slider changes the slopes of the intersecting lines at $b$ without moving $b$ itself, thereby modifying $a$ and $c$ automatically by altering the $x$-coordinates where the lines meet the $X$-axis. As intensity is the independent variable, these adjustments apply to $x$-coordinates.

\subsubsection{Fuzzy System}
For each pixel, the grayscale intensity is evaluated using the membership functions to obtain $\mu_{dark}$, $\mu_{gray}$, and $\mu_{bright}$. The aggregated output is:
\begin{equation}
I = \frac{v_d\,u_d(\text{input\_px}) + v_g\,u_g(\text{input\_px}) + v_b\,u_b(\text{input\_px})}{u_d(\text{input\_px}) + u_g(\text{input\_px}) + u_b(\text{input\_px})}
\end{equation}
where $v_d=0$, $v_g=127$, $v_b=255$. Based on rules, pixels with intensity $<80$ are black, $>140$ are white, and others are ambiguous, sent to the “Spatial Data Analysis” module.

\subsection{Spatial Data Analysis} The 'Adjusted Variogram' and 'Cumulative Squared Shift of Nodal Intensity' (CSSNI) are computed and normalized by $SSDLM_{\omega}$ for each pixel based on its $5 \times 5$ neighborhood. Pixels are first classified by the fuzzy system (intensity $<80$ as black, $>140$ as white), with ambiguous pixels subjected to spatial data analysis. Moran's I ensures pixels in disordered neighborhoods are not misclassified as white, with this rule adjustable via the calibration window. While probabilistic approaches address pixel-level uncertainty \cite{Lugagne2020}, our method provides a complementary deterministic solution combining fuzzy logic and local spatial descriptors.

\subsection{Final Masking} 
The final masking module resolves ambiguous pixels by analyzing intensity and spatial contrast in RGB channels. For mid-range fuzzy intensities, the normalized adjusted variogram within a $5 \times 5$ neighborhood is evaluated: $v_p^{\text{norm}} = \frac{\gamma(\mathcal{N}_p)}{\sigma(\mathcal{N}_p)}$. Pixels below the "NAV Threshold" (0–10, Fig.~\ref{fig:cal_denois}) undergo this analysis. Moran’s $I_p$ is computed over the same patch ("Randomness Threshold," $-1$ to $+1$) to detect significant spatial structure, and a heuristic rule based on channel intensity and contrast is applied.

The rule considers green intensity $G_p$ relative to red and blue ($G_p < R_p$ and $G_p < B_p$), and the contrast ordering $d_R > d_G > d_B$, where $d_*$ is the normalized CSSNI per channel. Pixels meeting both conditions are classified as background; otherwise, they remain foreground.

\begin{figure}[htbp!] 
  \centering  \includegraphics[width=.48 \textwidth]{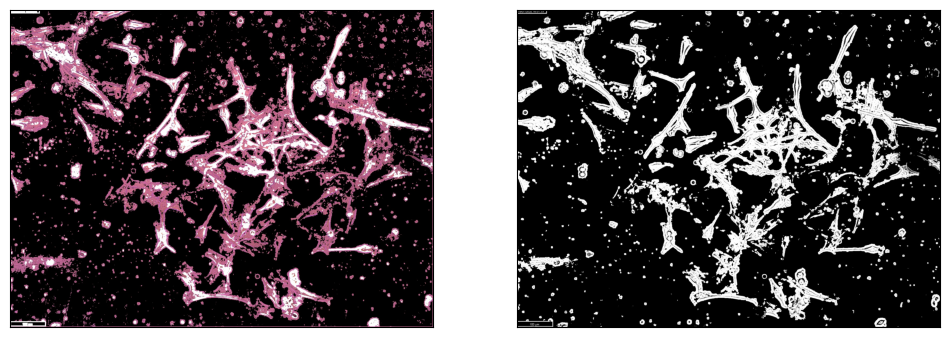}
  \caption{Image after masking based on lower bound and upper bound where pink color denotes the uncertainty regions (left), Segmented image with Noise (Right)}
  \label{fig:after_mask}
\end{figure}

\subsection{Hyperparameter \& Post-Segmentation Denoising:} Once the output is generated in the step called "Final Masking", Hyperparameter may be tuned for refining the output or denoising the output (if required) and replay the segmentation loop the desired result. Fig.~\ref{fig:cal_res}  demonstrates two different example outputs generated by two types of tuning of hyperparameter for visual impact.

\begin{figure}[htbp!]
  \centering
  \includegraphics[width=.48\textwidth]{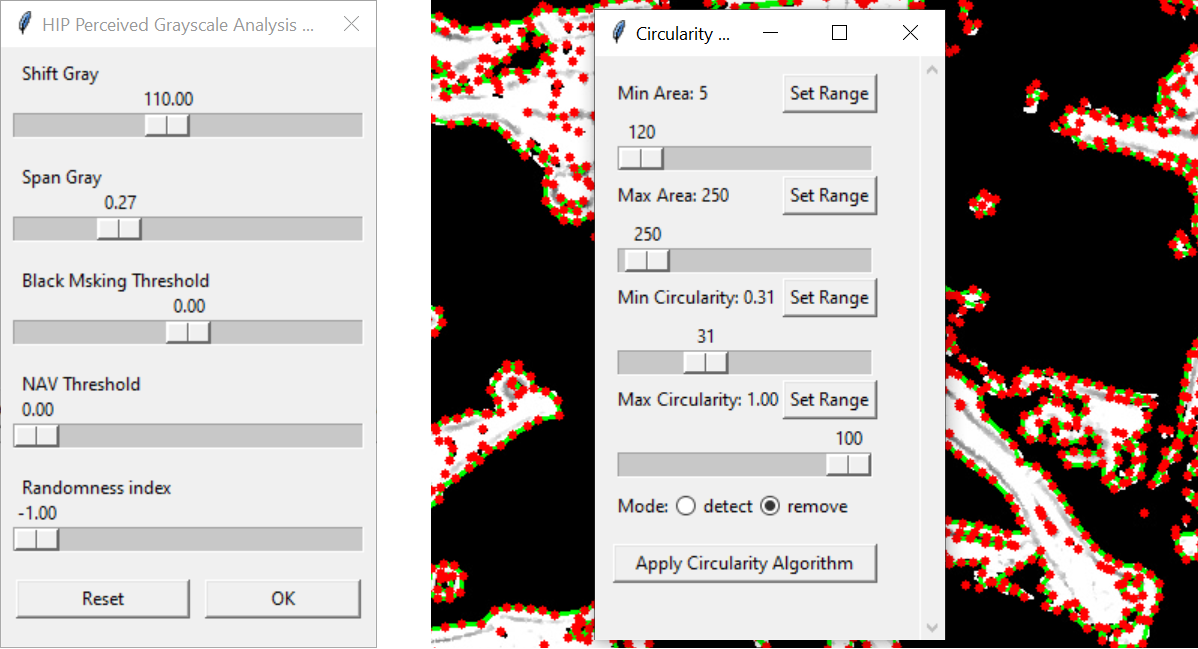}
  \caption{Left: Hyperparameter-Tuning Window, Right: Post-Segmentation Denoising Window}
  \label{fig:cal_denois}
\end{figure}

Post-Segmentation Denoising for (Dataset-1) begins with contour filling and morphological erosion of segmented objects
(applicable for all datasets), followed by removing protein blobs based on circularity (isoperimetric quotient) and specified area thresholds, discarding blobs outside these bounds. In this regards, areas below 100 pixels are filled. Two circularity filters then remove objects: the first targets small objects (area 5--293, any circularity), and the second targets larger, irregular objects (area 253--1800, circularity $<$ 0.31). A median blur (kernel size 5) is applied last (for all three datasets) to suppress salt-and-pepper noise. Fig.~\ref{fig:cal_denois} illustrates the denoising GUI. The denoising profile varies with imaging modality and cell type; in this study, two profiles were used for 3,178 images:

\begin{figure}[htbp!]
  \centering
  \includegraphics[width=.48\textwidth]{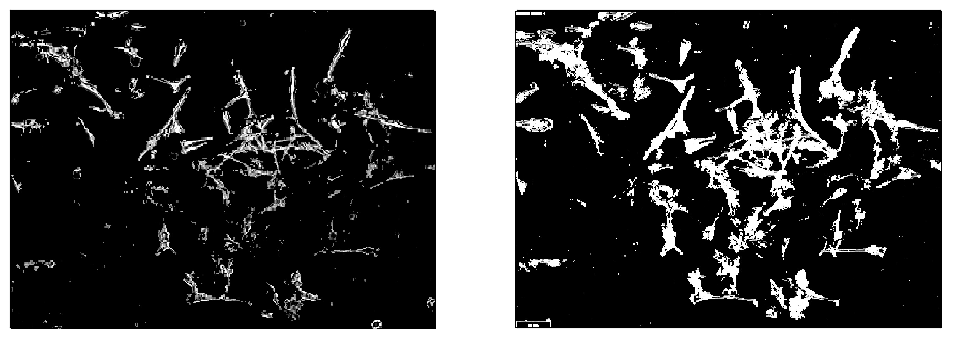}
  \caption{Two different outputs generated by two hyperparameter tunings}
  \label{fig:cal_res} 
\end{figure}

\textbf{Profile-1:} Lower bound (LB) = 4.23 (default); pipeline: fill-below-threshold (area 100), erosion (kernel 3), circularity filter (area 0--71, circularity 0--1, mode=remove), median blur (kernel 5).  

\textbf{Profile-2:} LB = 2.71; others are identical to Profile-1.

All images in the datasets 2 and 3 are processed with Profile-1, except for four categories of images in Dataset-2, where Profile-2 is used (Table~\ref{tab:image_group_f1_profile1}).

\subsection{Algorithm}
This section includes two algorithms based on which the unsupervised masking works: \\
Algorithm 1: Unsupervised Masking \\ 
\textbf{Input:} RGB image \( I \), patch size \( n \times n \)  \\
\textbf{Output:} Binary mask \( S \)  \\
01: Convert \( I \) to grayscale: \( I_{\text{gray}} = \frac{R + G + B}{3} \)  \\
02: Sample background pixels and compute \( \text{SSDLM} \) (Eq. 4)  \\
03: Set threshold \( L_B = \mu_{\text{SSDLM}} + 3\sigma_{\text{SSDLM}} \) \\
04: \textbf{for each} pixel \( p \in I_{\text{gray}} \) \textbf{do} \\
05: \quad Extract patch \( \mathcal{N}_p \in \mathbb{R}^{n \times n} \) around \( p \) \\
06: \quad Compute fuzzy value \( f_p \) via fuzzy rules (Eqs. 7–9)  \\
07: \quad \textbf{if} \( f_p < a \) \textbf{then} \\
08: \quad \quad Set \( S(p) \leftarrow 0 \) \\
09: \quad \textbf{else if} \( f_p > 140 \) \textbf{then} \\
10: \quad \quad Set \( S(p) \leftarrow 255 \) \\
11: \quad \textbf{else} \textit{/* ambiguous mid-range */} \\
12: \quad \quad Compute \( v_p^{\text{norm}} = \gamma(\mathcal{N}_p)/\sigma(\mathcal{N}_p) \) (Eq. 6) \\
13: \quad \quad \textbf{if} \( v_p^{\text{norm}} < \texttt{NAV Threshold} \) \textbf{then} \\
14: \quad \quad \quad Compute Moran’s \( I_p \) on \( \mathcal{N}_p \) \\
15: \quad \quad \quad \textbf{if} \( I_p < \texttt{Randomness Threshold} \) \textbf{then} Set \( S(p) \leftarrow 0 \) \\
16: \quad \quad \quad \textbf{else} \\
17: \quad \quad \quad \quad Compute \( d_R, d_G, d_B = \texttt{CSSNI} (\mathcal{N}_{R,G,B}) \)  \\
18: \quad \quad \quad \quad \textbf{if} \( G_p < 100 \land G_p < R_p \land G_p < B_p \) \\
19: \quad \quad \quad \quad \quad \textbf{and} \( d_R > d_G > d_B \) \textbf{then} Set \( S(p) \leftarrow 0 \) \\
20: \quad \quad \quad \quad \textbf{else} Set \( S(p) \leftarrow 255 \) \\
21: \quad \quad \quad \textbf{end if} \\
22: \quad \quad \quad Compute \( d_R, d_G, d_B = \texttt{CSSNI}(\mathcal{N}_{R,G,B}) \) \\
23: \quad \quad \quad \textbf{if} \( G_p < 100 \land G_p < R_p \land G_p < B_p \) \\
24: \quad \quad \quad \quad \textbf{and} \( d_R > d_G > d_B \) \textbf{then} Set \( S(p) \leftarrow 0 \) \\
25: \quad \quad \quad \textbf{else} Set \( S(p) \leftarrow 255 \) \\
26: \quad \quad \textbf{end if} \\
27: \quad \textbf{end if} \\
28: \textbf{end for} \\
29: \textbf{if} output \( S \) requires refinement \textbf{then} \\
30: \quad Apply postprocessing (e.g., denoising) \\
31: \textbf{else} calibrate thresholds and repeat steps 03–29 

\subsection{Model Evaluation}
To evaluate segmentation performance, we used a comprehensive set of standard metrics: Intersection over Union (IoU), pixel-wise accuracy, precision, recall and F1 score. Definitions and detailed formulations are standard in medical image analysis literature \cite{metrics_survey}, and thus omitted here for brevity.

\subsection{Statistical Analysis}
Performance differences between the proposed framework and baseline models were evaluated using non-parametric statistical testing. 
The Wilcoxon signed-rank test was applied to per-image Intersection over Union (IoU) and F1-scores to determine whether observed improvements were statistically significant, with a significance threshold of $p < 0.05$. 
To assess reliability of expert evaluation, inter-rater agreement between biologists and between each model and expert annotations was quantified using Cohen’s Kappa coefficient ($\kappa$), where values above 0.75 indicate substantial agreement. 
All statistical analyses were conducted at the per-image level to ensure fair and consistent comparisons across datasets.

\section{Results}

\subsection{Dataset-1 (unstained myoblast C2C12 cell images, n = 10, with bright-field modality. The dimension of each image is $1920 \times 1440$.)}

\begin{table}[!htbp]
\centering
\caption{Model Performance Comparison (Averages)}
\label{tab:model_avg_performance}
\resizebox{0.45\textwidth}{!}{
\begin{tabular}{lccccc}
\toprule
Model & IoU & Accuracy & Precision & Recall & F1 Score \\
\midrule
Our Model & 0.431 & 0.871 & 0.531 & 0.726 & 0.601 \\
StarDist  & 0.087 & 0.672 & 0.130 & 0.267 & 0.172 \\
Cellpose  & 0.130 & 0.865 & 0.358 & 0.164 & 0.205 \\
\bottomrule
\end{tabular}
}
\end{table}

\begin{table}[!htbp]
\centering
\caption{Model Performance Comparison for 10 images}
\label{tab:10_img_score}
\resizebox{0.48\textwidth}{!}{
\begin{tabular}{ccccccc}
\toprule
Img & Model & IoU & Accuracy & Precision & Recall & F1 Score \\
\midrule
01 & Our Model & 0.39 & 0.83 & 0.54 & 0.58 & 0.56 \\
01 & Stardist & 0.16 & 0.65 & 0.23 & 0.36 & 0.28 \\
01 & Cellpose & 0.09 & 0.81 & 0.40 & 0.11 & 0.17 \\
02 & Our Model & 0.43 & 0.95 & 0.50 & 0.75 & 0.60 \\
02 & Stardist & 0.14 & 0.87 & 0.18 & 0.41 & 0.25 \\
02 & Cellpose & $<$.01 & 0.94 & 0.04 & $<$.01 & $<$.01 \\
03 & Our Model & 0.47 & 0.88 & 0.57 & 0.74 & 0.64 \\
03 & Stardist & 0.09 & 0.82 & 0.13 & 0.25 & 0.17 \\
03 & Cellpose & 0.17 & 0.91 & 0.37 & 0.24 & 0.29 \\
04 & Our Model & 0.36 & 0.96 & 0.50 & 0.56 & 0.52 \\
04 & Stardist & 0.03 & 0.86 & 0.12 & 0.39 & 0.19 \\
04 & Cellpose & 0.03 & 0.95 & 0.18 & 0.03 & 0.05 \\
05 & Our Model & 0.43 & 0.89 & 0.53 & 0.70 & 0.61 \\
05 & Stardist & 0.05 & 0.56 & 0.06 & 0.20 & 0.09 \\
05 & Cellpose & 0.09 & 0.88 & 0.42 & 0.10 & 0.16 \\
06 & Our Model & 0.38 & 0.82 & 0.40 & 0.87 & 0.55 \\
06 & Stardist & 0.06 & 0.64 & 0.08 & 0.19 & 0.12 \\
06 & Cellpose & 0.06 & 0.85 & 0.24 & 0.08 & 0.11 \\
07 & Our Model & 0.39 & 0.82 & 0.40 & 0.91 & 0.56 \\
07 & Stardist & 0.05 & 0.65 & 0.07 & 0.15 & 0.09 \\
07 & Cellpose & 0.33 & 0.88 & 0.55 & 0.46 & 0.50 \\
08 & Our Model & 0.43 & 0.87 & 0.50 & 0.77 & 0.61 \\
08 & Stardist & 0.09 & 0.68 & 0.13 & 0.24 & 0.17 \\
08 & Cellpose & 0.01 & 0.86 & 0.07 & 0.01 & 0.01 \\
09 & Our Model & 0.55 & 0.91 & 0.64 & 0.79 & 0.71 \\
09 & Stardist & 0.06 & 0.57 & 0.08 & 0.22 & 0.12 \\
09 & Cellpose & 0.43 & 0.91 & 0.73 & 0.51 & 0.60 \\
10 & Our Model & 0.48 & 0.78 & 0.73 & 0.59 & 0.65 \\
10 & Stardist & 0.14 & 0.42 & 0.22 & 0.26 & 0.24 \\
10 & Cellpose & 0.09 & 0.66 & 0.58 & 0.10 & 0.16 \\
\bottomrule
\end{tabular}
} 
\end{table}

The proposed method achieved an average Intersection over Union (IoU) of 0.431, accuracy of 0.871, precision of 0.531, recall of 0.726, and F1 score of 0.601 (Table~\ref{tab:model_avg_performance}). 
By comparison, Cellpose achieved IoU 0.130 and StarDist 0.087. 
Per-image performance is listed in Table~\ref{tab:10_img_score}.

Statistical testing confirmed significance: the Wilcoxon signed-rank test (Table~\ref{tab:wilcoxon}) showed all p-values $<$ 0.01 when comparing our model against both Cellpose and StarDist. 

\begin{figure}[!htbp] 
  \centering  \includegraphics[width=.48 \textwidth]{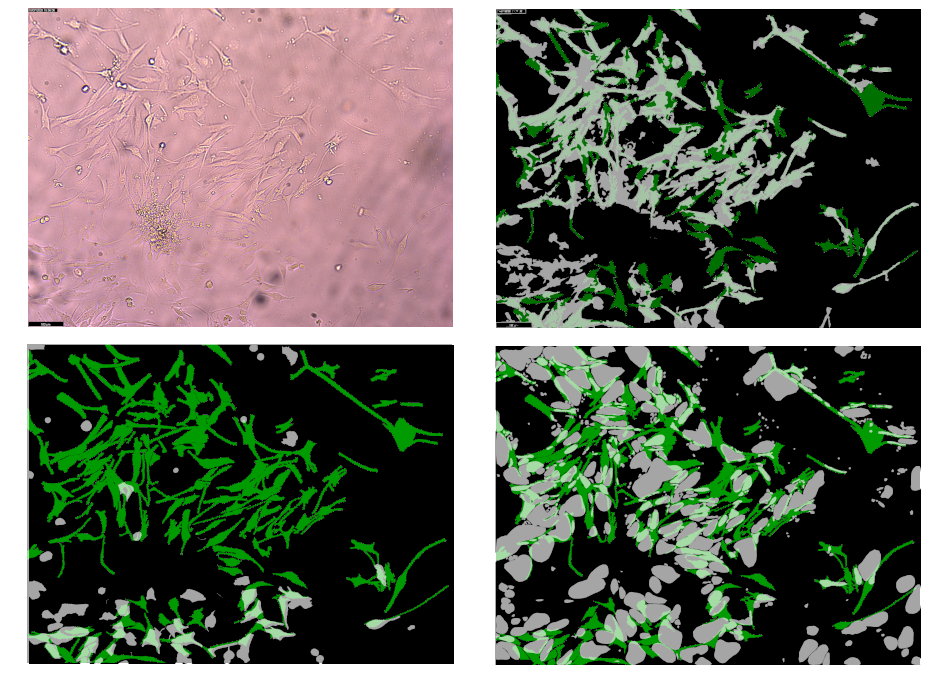}
  \caption{Top-left: original image. Top-right: our model. Bottom-left: Cellpose. Bottom-right: StarDist. The green pixels are the true positives which are not detected by a particular model. The bright white pixels are true positives which are detected by a model too. The off-white pixels are false positives detected by a model\\}
  \label{fig:comparison1}
\end{figure}

\begin{figure}[H] 
   \centering  \includegraphics[width=.48 \textwidth]{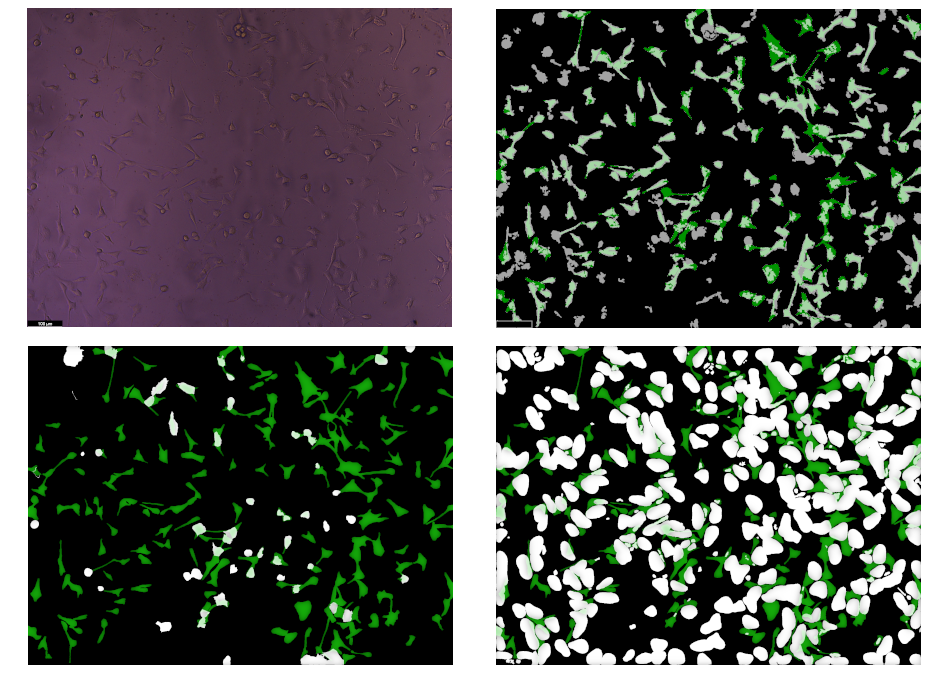}
  \centering  \includegraphics[width=.48 \textwidth]{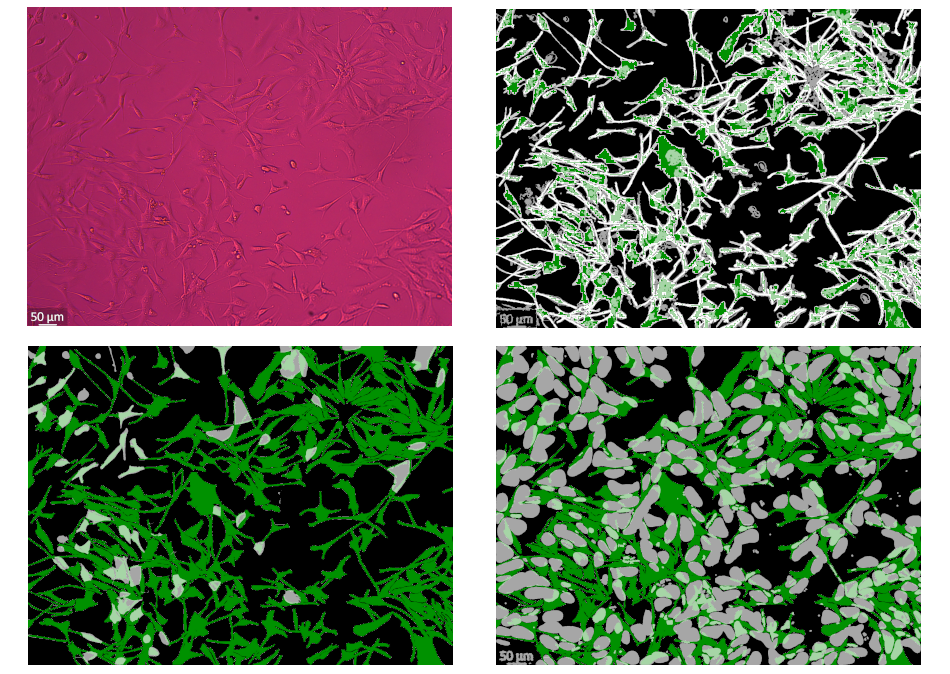}
  \centering  \includegraphics[width=.48 \textwidth]{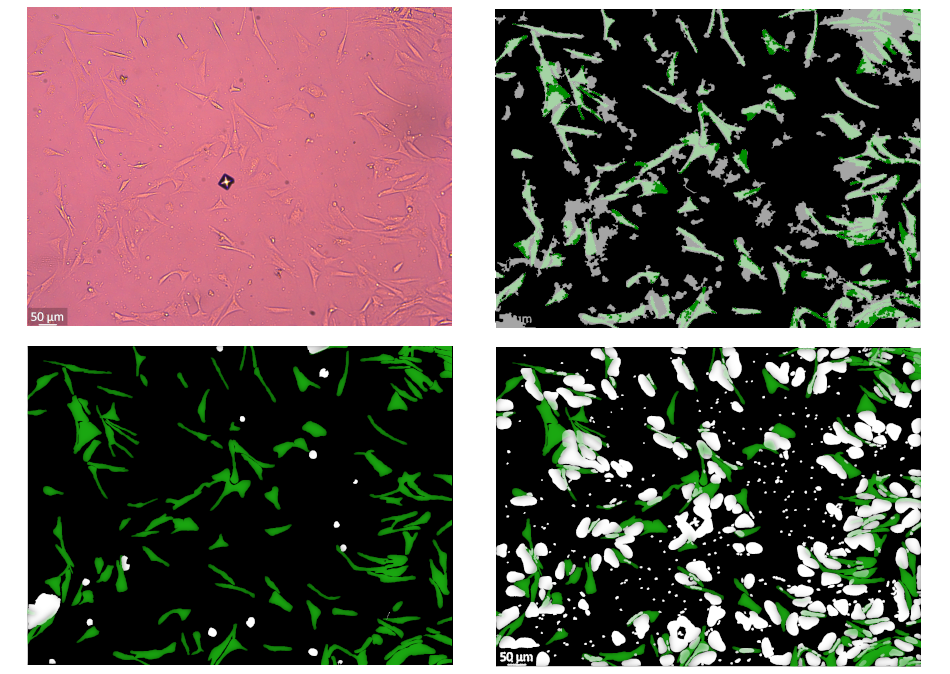}
  \caption{3 more images + segmentations (ours vs. SOTA), ordered as Fig. 5. Full results for Dataset-1 are given in the location specified in abstract.}
  \label{fig:comparison2}
\end{figure}

Expert evaluation indicated substantial agreement between biologists and the proposed model (Cohen’s $\kappa = 0.781$ and $0.754$ vs experts, compared to $0.421$ for Cellpose and $0.351$ for StarDist; Table~\ref{tab:expert_validation}). Representative segmentations are shown in Figures~\ref{fig:comparison1} and \ref{fig:comparison2}.

\begin{table}[!htpb]
\centering
\caption{Wilcoxon Signed-Rank Test Results}
\label{tab:wilcoxon}
\resizebox{0.32\textwidth}{!}{
\begin{tabular}{lcc}
\toprule
\\textbf{Comparison} & \textbf{Metric} & \textbf{p-value} \\
\midrule
Our Model vs. Cellpose & IoU & 0.004 \\
Our Model vs. StarDist & IoU & 0.002 \\
Our Model vs. Cellpose & F1-score & 0.002 \\
Our Model vs. StarDist & F1-score & 0.005 \\
\bottomrule
\end{tabular}
}
\end{table}

\begin{table}[!htpb]
\centering
\caption{Cohen's Kappa Agreement Scores}
\label{tab:expert_validation}
\resizebox{0.25\textwidth}{!}{
\begin{tabular}{lc}
\toprule
\textbf{Comparison} & \textbf{Kappa ($\kappa$)} \\
\midrule
Expert 1 vs. Expert 2         & 0.81 \\
Expert 1 vs. Our Model        & 0.78 \\
Expert 2 vs. Our Model        & 0.75 \\
Expert vs. Cellpose           & 0.42 \\
Expert vs. StarDist           & 0.35 \\
\bottomrule
\end{tabular}
}
\end{table}

\begin{table}[H]
\centering
\caption{Image group statistics with mean F1 score and standard deviation for Profile-1}
\label{tab:image_group_f1_profile1}
\resizebox{0.41\textwidth}{!}{
\begin{tabular}{lrrrl}
\hline
\textbf{Image Group} & \textbf{Count} & \textbf{Mean F1} & \textbf{Std} & \textbf{Profile} \\
\hline
MCF7\_Phase\_F4 & 152 & 0.892839 & 0.046123 & Profile-1 \\
SkBr3\_Phase\_E3 & 151 & 0.883631 & 0.019668 & Profile-1 \\
SkBr3\_Phase\_H3 & 151 & 0.883416 & 0.023487 & Profile-1 \\
SkBr3\_Phase\_F3 & 146 & 0.879517 & 0.028173 & Profile-1 \\
MCF7\_Phase\_E4 & 157 & 0.874425 & 0.052069 & Profile-1 \\
MCF7\_Phase\_G4 & 160 & 0.866882 & 0.067715 & Profile-1 \\
A172\_Phase\_B7 & 128 & 0.845009 & 0.043283 & Profile-1 \\
A172\_Phase\_A7 & 129 & 0.842233 & 0.046283 & Profile-1 \\
A172\_Phase\_D7 & 128 & 0.826452 & 0.045234 & Profile-1 \\
BV2\_Phase\_D4 & 123 & 0.809159 & 0.066581 & Profile-1 \\
BT474\_Phase\_B3 & 147 & 0.808844 & 0.056174 & Profile-1 \\
BT474\_Phase\_C3 & 140 & 0.808419 & 0.057977 & Profile-1 \\
BV2\_Phase\_C4 & 128 & 0.805938 & 0.065642 & Profile-1 \\
BT474\_Phase\_A3 & 141 & 0.797301 & 0.063672 & Profile-1 \\
BV2\_Phase\_B4 & 133 & 0.795488 & 0.093175 & Profile-1 \\
SHSY5Y\_Phase\_D10 & 146 & 0.791682 & 0.040150 & Profile-1 \\
SHSY5Y\_Phase\_B10 & 156 & 0.788556 & 0.045465 & Profile-1 \\
SHSY5Y\_Phase\_C10 & 146 & 0.784836 & 0.044050 & Profile-1 \\
Huh7\_Phase\_A10 & 174 & 0.765936 & 0.069446 & Profile-2 \\
Huh7\_Phase\_A11 & 176 & 0.733752 & 0.096981 & Profile-2 \\
SKOV3\_Phase\_H4 & 139 & 0.720993 & 0.060473 & Profile-2 \\
SKOV3\_Phase\_G4 & 127 & 0.708126 & 0.077193 & Profile-2 \\
\hline
\end{tabular}
}
\end{table}

\subsection{Dataset-2 (LIVECell, n = 3178)}
The framework achieved mean Dice score of 0.8144 $\pm$ 0.0786, IoU of 0.6940 $\pm$ 0.1069, accuracy of 0.8720 $\pm$ 0.0961, precision of 0.8034 $\pm$ 0.1125, and recall of 0.8512 $\pm$ 0.1246 (Table~\ref{tab:metrics2}). 
Structural similarity (SSIM) was 0.5029 $\pm$ 0.2458, and average Hausdorff distance was 57.2873 $\pm$ 34.9202. 
Group-wise statistics are reported in Table~\ref{tab:image_group_f1_profile1}. 

\begin{figure}[h]
    \centering
    \begin{minipage}{0.23\textwidth}
        \centering
        \includegraphics[width=\linewidth]{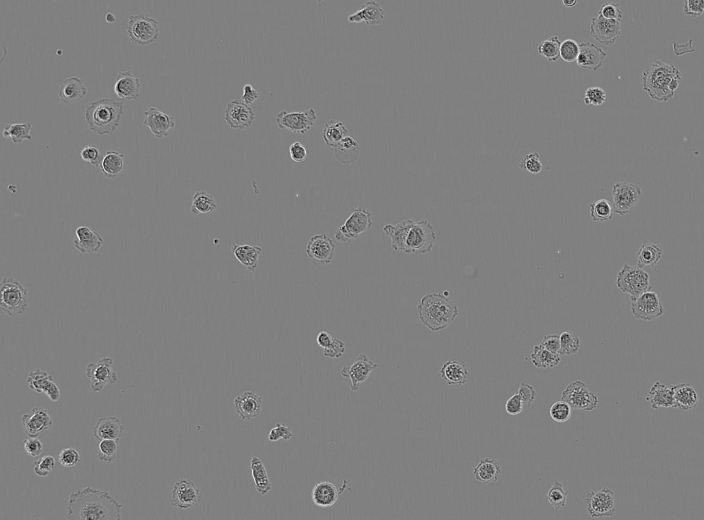}
    \end{minipage}
    \hfill
    \begin{minipage}{0.23\textwidth}
        \centering
        \includegraphics[width=\linewidth]{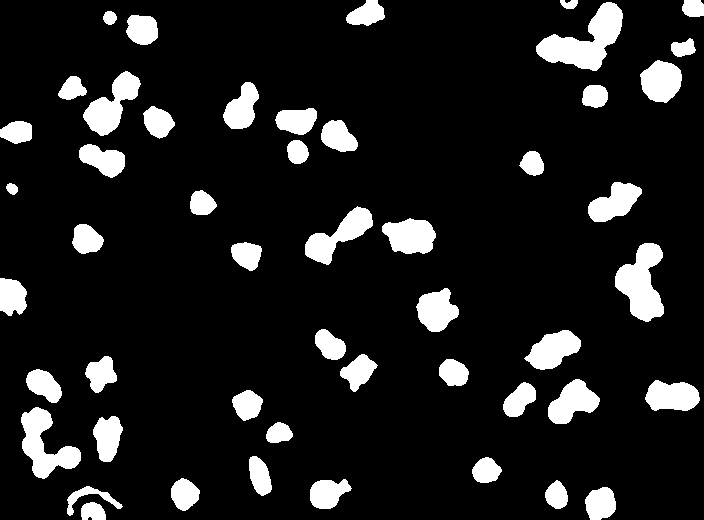}
    \end{minipage}
    
    \vspace{0.5cm} 
    
    \begin{minipage}{0.23\textwidth}
        \centering
        \includegraphics[width=\linewidth]{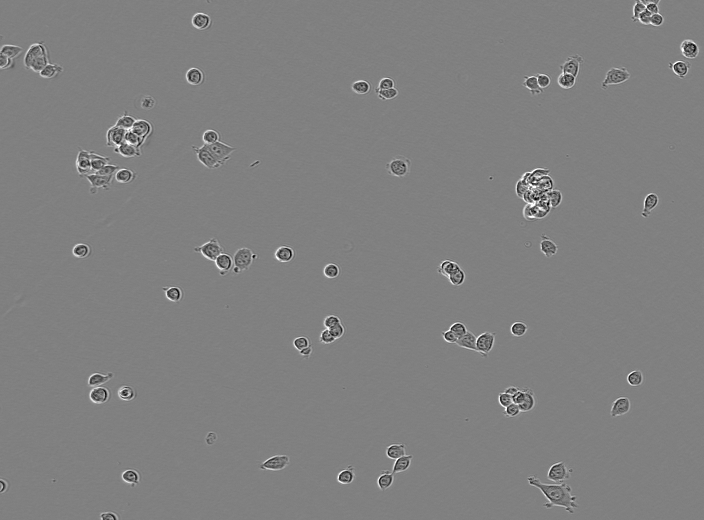}
    \end{minipage}
    \hfill
    \begin{minipage}{0.23\textwidth}
        \centering
        \includegraphics[width=\linewidth]{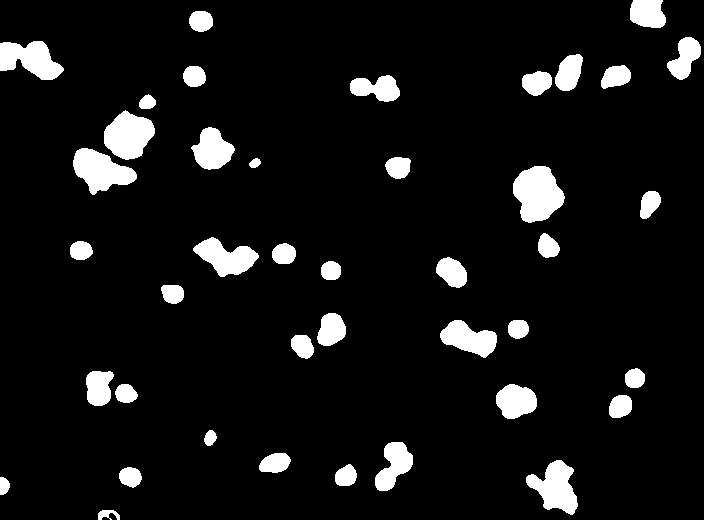}
    \end{minipage}
    
    \caption{Two instances with segmented results taken from the Dataset-2 (LIVECell Public Dataset). (Left) Original image, (Right) Output. Full results for Dataset-2 are given in the location specified in abstract}
    \label{fig:datafig2}
\end{figure}

Most cell categories performed with Profile-1, while four categories (Huh7\_Phase\_A10, Huh7\_Phase\_A11, SKOV3\_Phase\_G4, SKOV3\_Phase\_H4) required Profile-2, yielding mean F1 scores of 0.6689 $\pm$ 0.0659, 0.6415 $\pm$ 0.0956, 0.4742 $\pm$ 0.0817, and 0.5082 $\pm$ 0.0729 respectively. 
Representative results are shown in Figure~\ref{fig:datafig2}.

During the analysis, the following eight images were excluded as their masks were found to be corrupted:
 \texttt{A172\_Phase\_A7\_1\_01d04h00m\_3.png},  \texttt{A172\_Phase\_D7\_1\_01d20h00m\_1.png}, 
 \texttt{BT474\_Phase\_B3\_1\_03d00h00m\_3.png}, 
 \texttt{BV2\_Phase\_C4\_1\_01d16h00m\_3.png}, 
 \texttt{BV2\_Phase\_D4\_1\_00d12h00m\_2.png},  
 \texttt{Huh7\_Phase\_A11\_1\_00d04h00m\_3.png},  
 \texttt{Huh7\_Phase\_A11\_1\_00d04h00m\_4.png},     
 \texttt{SHSY5Y\_Phase\_D10\_1\_01d16h00m\_4.png}

\begin{table}[h]
\centering
\caption{Dataset-2: Segmentation performance metrics}
\label{tab:metrics2}
\resizebox{0.48\textwidth}{!}{%
\begin{tabular}{lcccccccc}
\hline
\textbf{ } & \textbf{Dice} & \textbf{IoU} & \textbf{Accuracy} & \textbf{Precision} & \textbf{Recall} & \textbf{F1 Score} & \textbf{SSIM} & \textbf{Hausdorff} \\
\hline
Mean & 0.814431 & 0.694019 & 0.872015 & 0.803413 & 0.851245 & 0.814431 & 0.502930 & 57.287395 \\
Std  & 0.078670 & 0.106993 & 0.096104 & 0.112562 & 0.124639 & 0.078670 & 0.245851 & 34.920233 \\
\hline
\end{tabular}
}
\end{table}

\subsection{Dataset-3 (Controllable Laser Trace)}
\begin{figure}[H]
    \centering
    \begin{minipage}{0.23\textwidth}
        \centering
        \includegraphics[width=\linewidth]{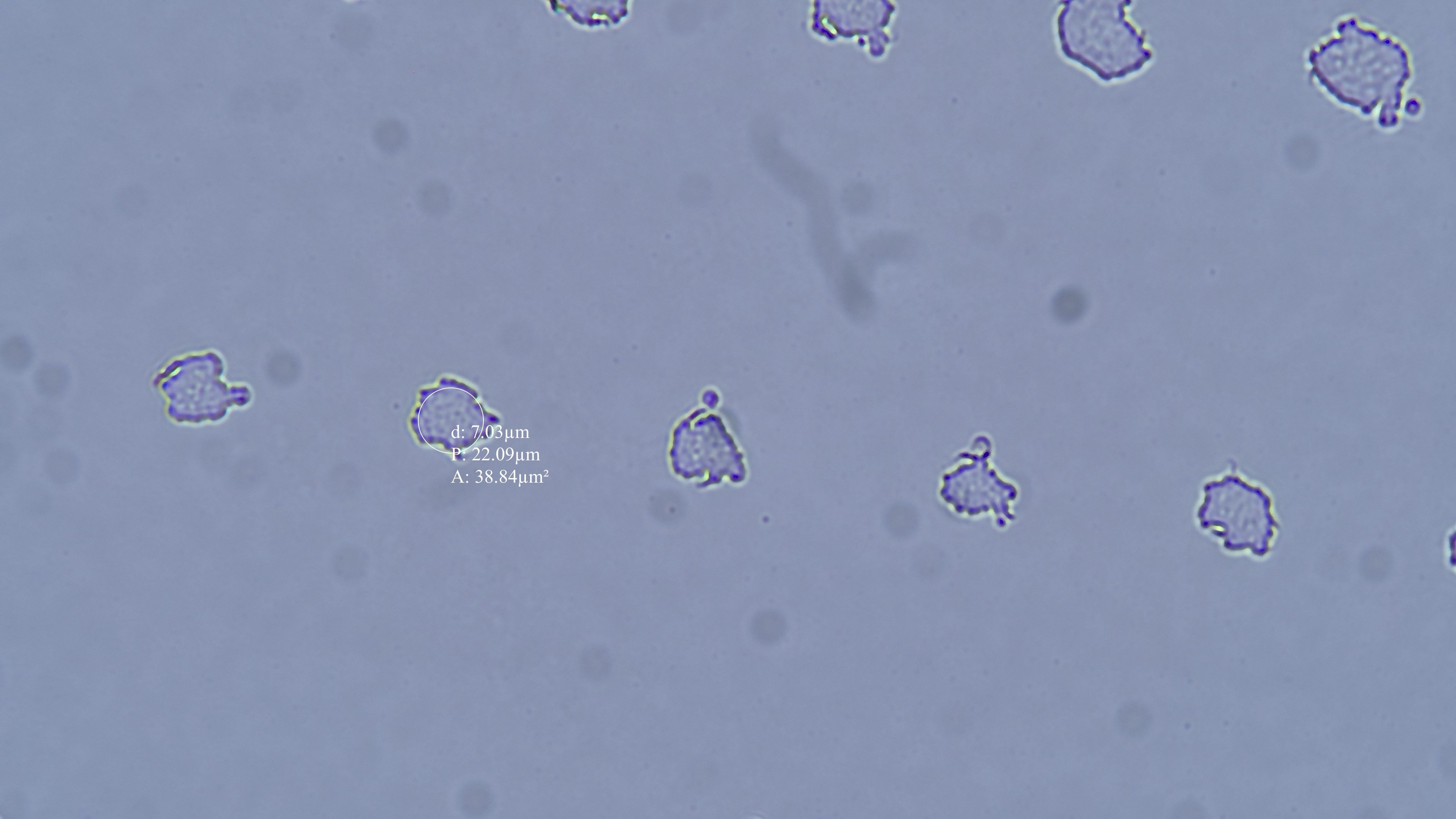} 
    \end{minipage}
    \hfill
    \begin{minipage}{0.23\textwidth}
        \centering
        \includegraphics[width=\linewidth]{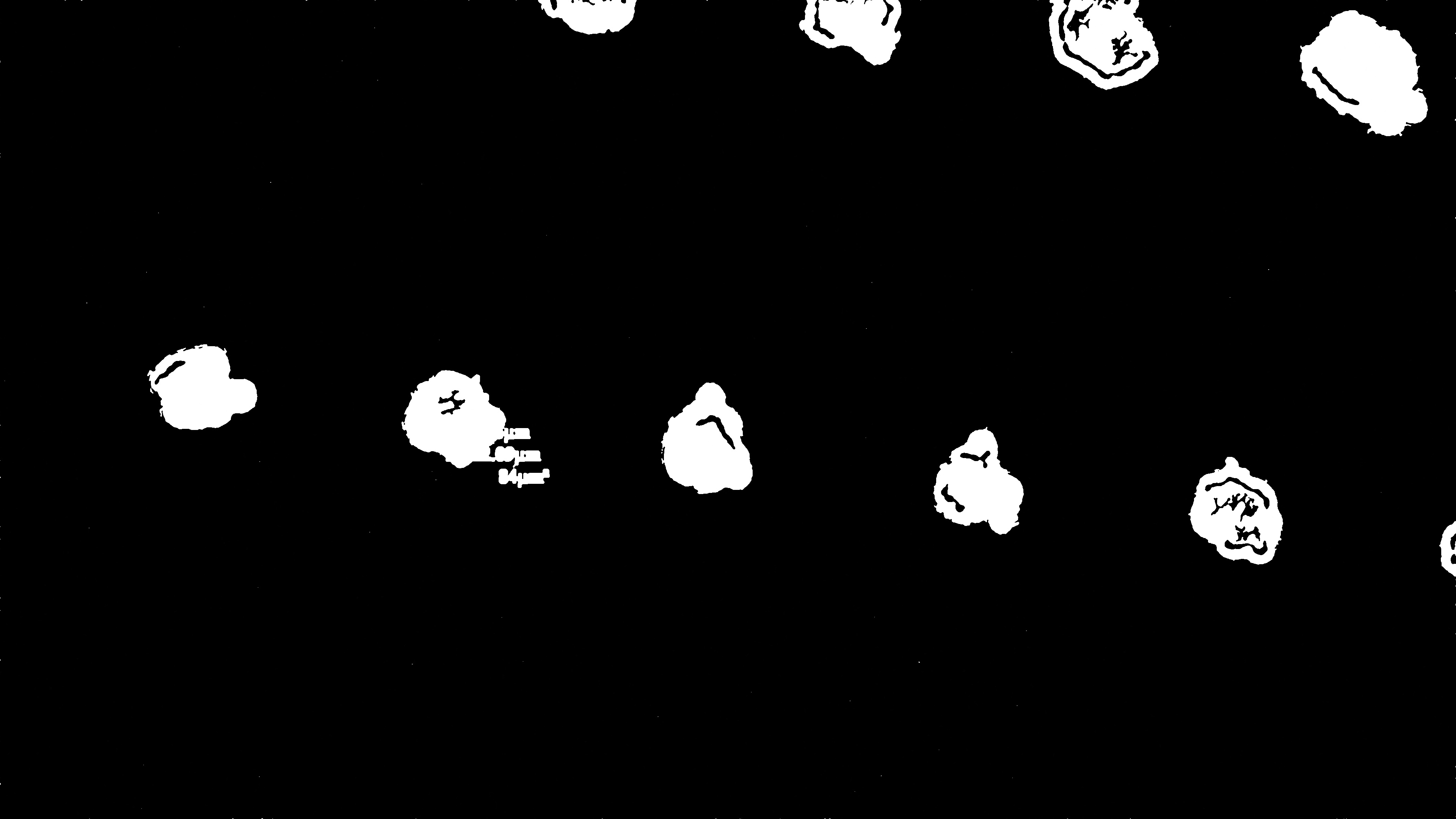} 
      \end{minipage}
    \caption{Segementation Result from Dataset-3 (Trace of laser); (left) Original Image, (Right) Output.}
    \label{fig:datafig3}
\end{figure}

The framework maintained consistent segmentation quality (with F1-score $0.83 - 0.90$) on simulated laser-trace images, as shown in Figure~\ref{fig:datafig3}.

\section{Discussion}

The results demonstrate that the proposed framework consistently outperforms Cellpose and StarDist on unstained brightfield images, achieving up to 48\% higher IoU on Dataset-1.

All Wilcoxon signed-rank tests comparing the proposed method with Cellpose and StarDist yielded $p$-values below 0.05, confirming that the observed improvements in IoU and F1-scores were statistically significant rather than random variation. 
Furthermore, Cohen’s Kappa analysis indicated substantial agreement ($\kappa > 0.75$) between expert annotations and the proposed framework, reinforcing the reliability of the segmentation outputs in practical biomedical applications related to hard challenge -- low-contrast microscopy.

The findings validate the central concept of the \textit{Homogeneous Image Plane}. 
In microscopy, objects appear heterogeneous due to scattering and texture, while the background remains nearly homogeneous. 
By explicitly exploiting this property with spatial statistics and fuzzy inference, the framework separates object heterogeneity from background homogeneity.

Compared with deep learning approaches, our method has three key advantages:
\begin{enumerate}
    \item \textbf{Training-free and annotation-free} --- no labeled data or retraining needed.  
    \item \textbf{Robustness across modalities} --- three profiles sufficed for 3200+ images from diverse cell lines. Calibration is required only when imaging modality or specimen type changes substantially.  
    \item \textbf{Accessibility} --- runs efficiently on CPU hardware, usable via both GUI / scripts, and suited to laboratories without GPUs or annotation resources.  
\end{enumerate}

A limitation of the study is the relatively small size of Dataset-1. 
However, this was addressed by validation on the large LIVECell dataset ($n > 3000$). 
Precision values were somewhat lower than Cellpose, reflecting a conservative design that prioritizes recall and structural fidelity --- a trade-off beneficial in regenerative medicine, where false negatives are more critical than false positives.

Future extensions include automating calibration profile selection, expanding validation to additional modalities, and integrating explainable AI for improved interpretability.

Overall, the framework introduces a novel theoretical foundation through the \textit{Homogeneous Image Plane} concept and demonstrates practical robustness across microscopy modalities, offering an alternative to data-hungry deep learning methods.

\section{Conclusion}

This method enables robust, unsupervised, training-free segmentation of bright-field microscopy images and is validated through standard performance metrics (Accuracy, Precision, Recall, IoU, F1-score), statistical significance testing, and expert visual assessment. While it allows calibration through an interactive interface, the segmentation process is largely automated and does not require annotated training data. This makes it well-suited for regenerative medicine applications, such as stem cell tracking, wound healing assays, and time-lapse live-cell imaging. The proposed approach supports fast, reproducible analysis in experimental and clinical microscopy workflows. Future directions encompass better precision and instance segmentation for overlapping cellular communities. Also, despite the promising results, our study has limitations pertaining to the size of the dataset and relatively low precision which would be addressed in future work.

\section*{Acknowledgments}
The authors would like to thank Svetlana Ulasevich and Andrei Buzykin for providing microscopy images and for fruitful
discussions of the domain area.

\section*{Funding}
The research was supported by ITMO University Research Projects in AI Initiative (RPAII) (project \#640103, Development of methods for automated processing and analysis of optical and atomic force microscopy images using machine learning techniques).

\section*{Declaration of Competing Interest}
The authors declare that they have no known competing financial interests or personal relationships that could have appeared to influence the work reported in this paper\footnote{This publication is a preprint and has not been peer-reviewed. It is made available under the selected license. Any unauthorized use or violation of the license terms is strictly prohibited and will be the responsibility of the individual or entity engaging in such use.}.

\section*{Ethics Statement}
This study did not involve human participants or animals.

\section*{Author Contributions}
S.D. designed the study, analyzed the data, implemented the
model, drafted the manuscript, and supervised the work. P.Z.
provided supervision and critical review of the manuscript. All
authors read and approved the final version.

\bibliographystyle{elsarticle-num}

\end{document}